\documentclass[prb,showpacs,twocolumn,byrevtex]{revtex4}
\usepackage{graphicx}
\usepackage{float}
\usepackage{bm}
\begin{document}

\bibliographystyle{apsrev}

\preprint{Draft version, not for distribution}

%
%
\title[coherence]{Superfluid density in overdoped cuprates: thin films versus bulk samples}

%
%
%
\author{S.V. Dordevic}
\email{dsasa@uakron.edu}%
\affiliation{Department of Physics, The University of Akron,
Akron, Ohio 44325, USA}%
\author{C.C. Homes}
\affiliation{Condensed Matter Physics and Materials Science Division,
Brookhaven National Laboratory, Upton, New York 11973, USA}%

\date{\today}

%
%
\begin{abstract} Recent study of overdoped La$_{2-x}$Sr$_x$CuO$_4$ cuprate superconductor
thin films by Bo\v{z}ovi\'{c} {\it et al.} has revealed several unexpected
findings, most notably the violation of the BCS description which
was believed to adequately describe overdoped cuprates.
In particular, it was found that the superfluid density in La$_{2-x}$Sr$_x$CuO$_4$ films
decreases on the overdoped side as a linear function of critical
temperature T$_c$, which was taken as evidence for the violation of the Homes' law.
We show explicitly that the law is indeed violated, and as the main
reason for violation we find that the superfluid
density in Bo\v{z}ovi\'{c}'s films is suppressed more strongly
than in bulk samples. Based on the existing literature data,
we show that the superfluid density in bulk cuprate samples
does not decrease with doping, but instead tends to saturate
on the overdoped side. The result is also supported by our recent
measurement of a heavily overdoped bulk La$_{2-x}$Sr$_x$CuO$_4$ sample. 
Moreover, this saturation of superfluid density might not be limited to cuprates,
as we find evidence for similar behavior in two pnictide 
superconductor families. We argue that quantum phase 
fluctuations play an important role in suppressing 
the superfluid density in thin films. 
\end{abstract}

%
%
%
%

\maketitle

\section{Introduction}

A recent finding by Bo\v{z}ovi\'{c} {\it et al.} \cite{bozovic16} that behavior of
cuprate superconductors on the overdoped side of their phase diagram
deviates strongly from the expected BCS behavior has generated a lot
of attention. Bo\v{z}ovi\'{c} {\it et al.} analyzed thousands of La$_{2-x}$Sr$_x$CuO$_4$
(LSCO) thin films, with thicknesses ranging from 0.66~nm to over 100~nm, 
and found that on the overdoped side the superfluid
density decreases as a linear function of superconducting critical temperature T$_c$,
and eventually goes to zero at the quantum critical point.
This is stark contrast with the Homes' law \cite{homes05}:

\begin{equation}
  \rho_s \propto T_c\, \sigma_{dc}
  \label{eq:homes}%
\end{equation}
where $\rho_s$ is the superfluid density and $\sigma_{dc}$ is dc conductivity just
above T$_c$. Eq.~\ref{eq:homes} predicts that when dc conductivity increases as 
a result of doping, the superfluid density should increase as well, assuming that 
the same percentage of charge carriers condenses. This is
in sharp contrast with the results of Bo\v{z}ovi\'{c} {\it et al.}

\begin{figure*}[t]
\vspace*{-1.0cm}%
\centerline{\includegraphics[width=6.5in]{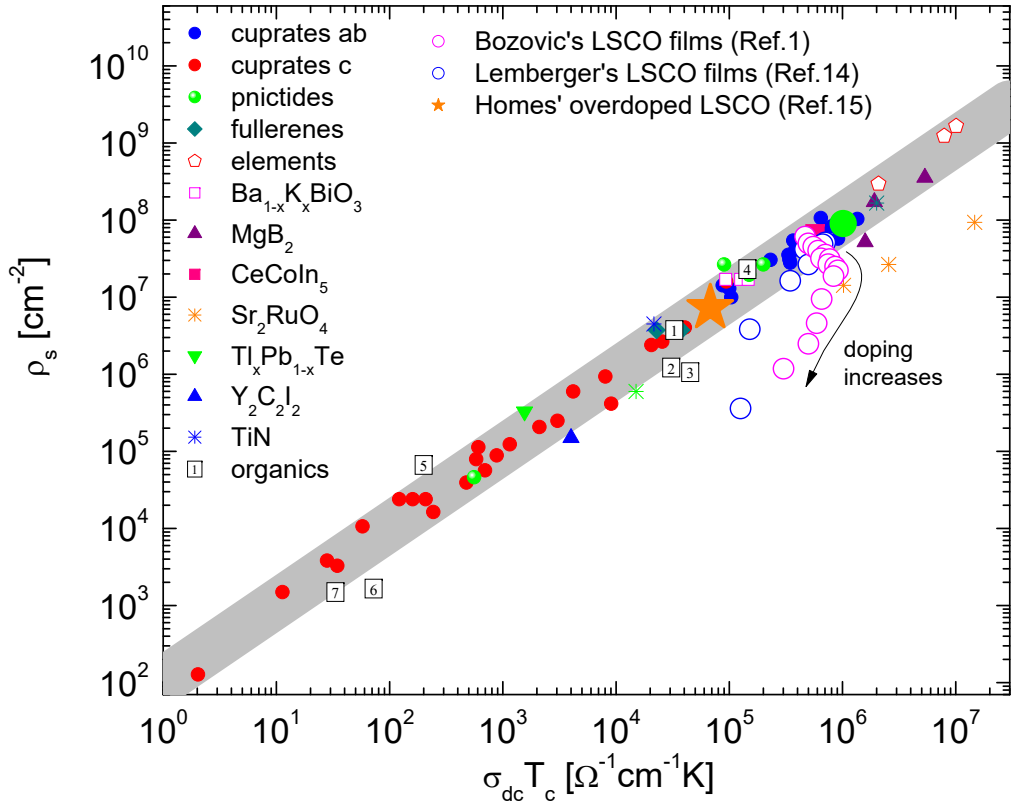}}%
\vspace*{-1.0cm}%
\caption{(Color online). Updated Homes' plot \cite{homes05,dordevic13} which
includes the data on Bo\v{z}ovi\'{c}'s LSCO films from
Ref.~\onlinecite{bozovic16}. We notice that Bo\v{z}ovi\'{c}'s data viloate
the scaling (Eq.~\ref{eq:homes}): the points progress perpendicular to the scaling
line as doping increases on the overdoped side of the phase diagram. The plot
also includes several overdoped LSCO films from  Lemberger's group
\cite{lemberger11}, which also violate the scaling. A big orange star 
represents our recent measurement on a heavily overdoped LSCO single 
crystal \cite{homes20}. Chemical formulas for all organic supercondcutors
can be found in Ref.~\onlinecite{dordevic13}.}
\vspace*{0.0cm}%
\label{fig:homes}
\end{figure*}

In Fig.~\ref{fig:homes} we display updated Homes plot \cite{homes05},
which includes all the previous data \cite{dordevic13}, along
with Bo\v{z}ovi\'{c}'s new data on LSCO films shown with open magenta circles.
The plot clearly shows that Bo\v{z}ovi\'{c}'s data does indeed
violate the Homes' law. The optimally doped film, as well as the films
close to that doping level are on the scaling line. However, as doping
increases on the overdoped side, the points move off the scaling line
and progress {\it perpendicular} to the line. At the highest doping levels,
the points take an additional downturn.


Kogan recently offered an explanation of Homes' scaling \cite{kogan13}, 
as well as the deviations from it. He argued that Homes' law
is a direct consequence of BCS theory and applies not only to dirty,
but also to moderately clean superconductors. Moreover, Kogan showed that
for clean superconductors, for which the ratio of superconducting
coherence length $\xi_0$ and the mean free path {\it l} is on the
order $\xi_0$/$l \sim$ 1 or smaller,
deviations from the scaling are expected. He predicted that
clean superconductors should move {\it below} the scaling line.
That behavior had indeed been observed previously\cite{dordevic13} in
Sr$_2$RuO$_4$ (three different samples
are shown with orange flakes in Fig.~\ref{fig:homes}),
which has been known to be a clean-limit superconductor ($\xi_0$/$l \ll$1).
Similarly, it was shown that elemental niobium in the clean limit (when recrystallized
in ultra high vacuum) also moves below the scaling line \cite{homes05b}.
Bo\v{z}ovi\'{c}'s LSCO films have been known to be ultra clean,
with a mean free path $l$~$\gtrsim$~4~$\mu$m (Ref.~\onlinecite{bozovic16}), 
exceeding their in-plane
coherence length (on the order of 20--30 $\AA$, Ref.~\onlinecite{wen03})
by at least two orders of magnitude.
Therefore, it is not surprising that Bo\v{z}ovi\'{c}'s films violate the
Homes' law and move {\it below} the scaling line.
However, as we show below, there could be other reasons for the
violation of Homes' law.


\section{Superfluid density}

\begin{figure}[t]
\vspace*{-1.0cm}%
\centerline{\includegraphics[width=9.5cm]{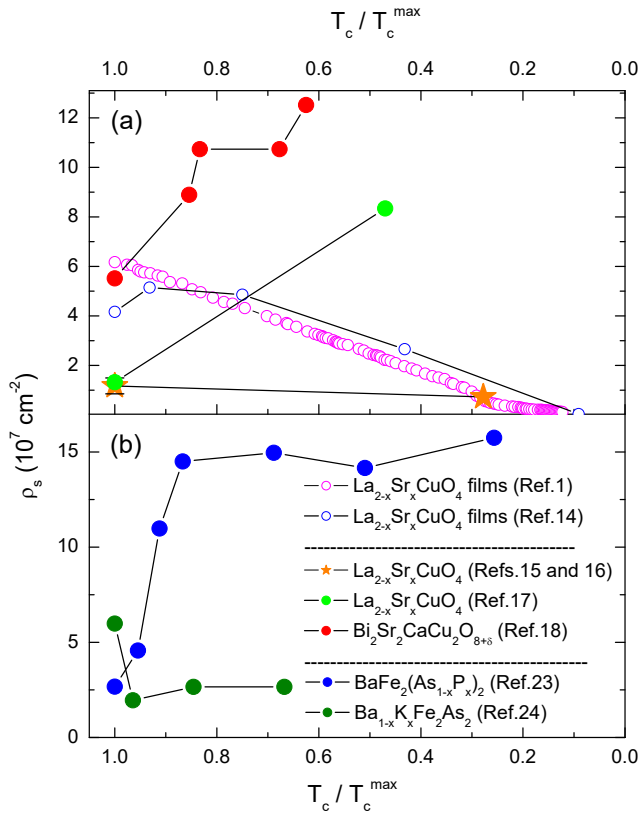}}%
\vspace*{-1.0cm}%
\caption{(Color online). (a) Superfluid density $\rho_s$ of LSCO and Bi2212
from several different measurements as a function of critical temperature,
normalized to the maximum value for a given family T$_c$/T$_c^{max}$
(Ref.\onlinecite{comm-Tcmax}). Only the overdoped side of the
phase diagram is shown and T$_c$/T$_c^{max}$ = 1 corresponds to optimal
doping. We notice that in films (Bo\v{z}ovi\'{c} \cite{bozovic16} and
Lemberger \cite{lemberger11} data displayed with open circles)
the superfluid density decreases with doping and eventually
goes to zero. On the other hand, in bulk samples (full circles) the superfluid
density tends to saturate \cite{tajima05,homes20} or even increase \cite{marel21}. 
In Bi2212 the superfluid density also 
increases with doping \cite{timusk07}. (b) Superfluid density $\rho_s$
of two pnictide families of superconductors \cite{matsuda12,almoalem18}. 
They are all bulk samples, and they all show saturation of 
superfluid density. }
\vspace*{0.0cm}%
\label{fig:overdoped}
\end{figure}

In order to explore the violation of Homes' law systematically
we have conducted an extensive literature search for the relevant experimental
parameters from Eq.~(\ref{eq:homes}) ($\rho_s$, $\sigma_{dc}$ and T$_c$).
Even though studies of the overdoped side are scarce, we have
collected enough data to unveil the trends.
Early reports of measurements performed on ceramic, sintered, polycrystalline or powder
samples were not considered here \cite{poly-comm}.
In Fig.~\ref{fig:overdoped}(a) we plot the superfluid density $\rho_s$ 
for two families of cuprate superconductors: LSCO 
\cite{bozovic16,lemberger11,tajima05,homes20,marel21} and 
Bi$_2$Sr$_2$CaCu$_2$O$_{8+\delta}$
(Bi2212) \cite{timusk07}. The superfluid density is shown as
a function of reduced critical temperature T$_c$/T$_{c, max}$, where
T$_{c, max}$ is the maximal critical temperature (i.e. optimal doping)
for a given family \cite{comm-Tcmax}. Only the overdoped side of
the phase diagram is shown, with T$_c$/T$_{c, max}$ = 1 being the
optimal doping and T$_c$ decreases as doping increases.
We point out that Fig.~\ref{fig:overdoped}(a) includes the data for both thin films 
(shown with open circles) and bulk single crystals (shown with full circles).
The plot also includes our recent infrared (IR) measurement \cite{homes20}
on a heavily overdoped bulk single crystal LSCO with T$_c$=~15~K, combined with
a previous IR measurement on optimally doped LSCO \cite{tajima05}.
We note that this heavily overdoped sample does not violate 
the Homes scaling (see Fig.~\ref{eq:homes}). 



As shown previously \cite{bozovic16} the superfluid density of 
Bo\v{z}ovi\'{c}'s films \cite{bozovic16} decreases as a linear
function of T$_c$, except at the highest dopings where the dependence
becomes parabolic. 
For comparison, Fig.~\ref{fig:overdoped}(a) also includes the results
on LSCO films from Lemberger's group \cite{lemberger11}. Even though
only several overdoped films were measured \cite{lemberger11},
they show very similar absolute values and doping dependence 
as Bo\v{z}ovi\'{c}'s films, and they also violate Homes' scaling
(see Fig.~\ref{fig:homes}). However,
the most striking finding revealed by Fig.~\ref{fig:overdoped}(a)
is that the superfluid density in bulk samples (full symbols) 
does not seem to decease with doping. In bulk single crystal LSCO samples
the superfluid density tends to saturate \cite{tajima05,homes20} 
or increase on the overdoped side \cite{marel21}.
The superfluid density in Bi2212 also increases 
with doping \cite{timusk07}. It is clear from Fig.~\ref{fig:overdoped}(a)
that the trend in the doping dependence is different for bulk samples and
thin films: the superfluid density in bulk samples, contrary to thin films,
does not decrease on the overdoped side.


\section{Pnictides}


The saturation of superfluid density observed in bulk samples
(Fig.~\ref{fig:overdoped}(a)) might not be limited to cuprates 
\cite{film-comm}. There is evidence that a similar effect is also present 
in at least two pnictide families. In Fig.~\ref{fig:overdoped}(b)
we show recent data on bulk single crystal pnictides
BaFe$_2$(As$_{1-x}$P$_x$)$_2$ \cite{matsuda12} and 
Ba$_{1-x}$K$_x$Fe$_2$As$_2$ \cite{almoalem18}. Both families reveal saturation 
of superfluid density on the overdoped side \cite{pnictide-comm}. 
It can be seen that in 
BaFe$_2$(As$_{1-x}$P$_x$)$_2$ the saturation persists up to very high
doping levels (T$_c$/T$_c^{max} \approx$ 0.25).
We also point out that pnictides are multiband systems with dramatically
different scattering rates associated with transport in different
bands, so they can effectively be superconductors that are in both
the clean and dirty limit at the same time \cite{homes15}.
Measurements of pnictide thin films are currently not available, 
but we hypothesize that on the overdoped side of the phase diagram, 
just like in the cuprates, their superfluid density might also be 
reduced compared to bulk samples.

\section{Normal state conductivity}

\begin{figure}[t]
\vspace*{-1.0cm}%
\centerline{\includegraphics[width=9.5cm]{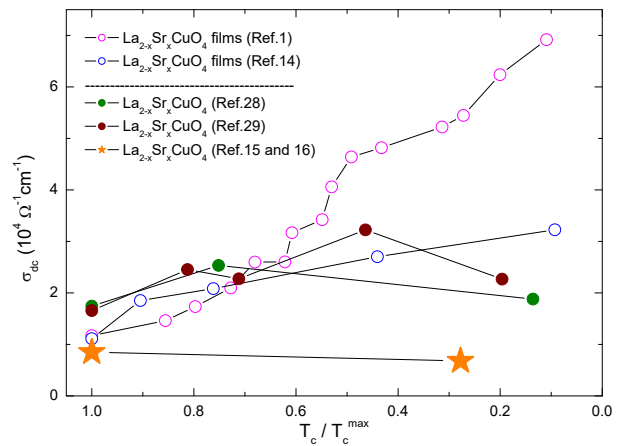}}%
\vspace*{-0.5cm}%
\caption{(Color online). Normal state conductivity at T$_c$, $\sigma_{dc}$, for several
LSCO bulk samples and films. $\sigma_{dc}$ is shown as a function of critical temperature,
normalized to the maximum value for a given family T$_c$/T$_c^{max}$
(Ref.\onlinecite{comm-Tcmax}). Only the overdoped side of the
phase diagram is shown and T$_c$/T$_c^{max}$ = 1 corresponds to optimal
doping. We notice that Bo\v{z}ovi\'{c} films \cite{bozovic16} have 
the strongest doping dependence, and at the highest doping levels
their values are several times greater compared to other films or
bulk samples. Lemberger films \cite{lemberger11} 
have significantly lower conductivity, even though they have 
comparable superfluid density (Fig.~\ref{fig:overdoped}(a)). 
On the other hand, bulk samples (full circles) display little
doping dependence of their $\sigma_{dc}$. The values of $\sigma_{dc}$
obtained from IR spectroscopy \cite{homes20,tajima05} 
(orange stars) are slightly lower.}
\vspace*{0.0cm}%
\label{fig:conductivity}
\end{figure}

In this section we analyze and compare the values of normal state 
conductivity for different LSCO samples. 
In Fig.~\ref{fig:conductivity} we display the values of dc conductivity 
just above T$_c$, $\sigma_{dc}$, from several different measurements
on LSCO. The plot includes both thin films (open circles) from Bo\v{z}ovi\'{c}
\cite{bozovic16} and Lemberger \cite{lemberger11} groups, as well as two
sets of measurements on bulk single crystals \cite{ando04,hussey09}
(full circles). We also plot the results of IR measurements 
on an optimally and a heavily overdoped LSCO single crystals \cite{tajima05,homes20}. 
Similar to the superfluid density (Fig.~\ref{fig:overdoped})
the conductivity is shown as a function of reduced critical temperature
T$_c$/T$_{c, max}$, and only on the overdoped side of the phase diagram.
It is immediately clear that Bo\v{z}ovi\'{c}'s films are superior in terms of their
conductivity. They show the strongest doping dependence, and 
at the highest doping levels (T$_c$/T$_c^{max} \approx$ 0.1)
their conductivity is approximately three times higher compared to all
other samples (bulk or film). We also note that overdoped bulk samples
might have issues with inhomogeneities \cite{homes20}, which can result 
in their conductivity being lower compared to films.


In spite of this finding, we argue that this high normal state conductivity 
of Bo\v{z}ovi\'{c}'s films is not the main reason
for the violation of Homes' law. Namely, Fig.\ref{fig:homes} also includes
the data for Lemberger films (open blue circles) which also violate the scaling,
i.e. as doping increases they move below the scaling line. Even though
their conductivity is several times smaller compared with Bo\v{z}ovi\'{c}'s 
films, their superfluid density is comparable and it
follows similar doping dependence (Fig.~\ref{fig:overdoped}(a)).  Therefore
we can conclude that the main reason for the Homes' law violation is the
suppression of superfluid density in thin films compared with bulk crystals.

\section{Data variability}


We have previously argued \cite{dordevic13} that for accurate scaling
the data for Eq.~\ref{eq:homes} ($\rho_s$, T$_c$ and $\sigma_{dc}$) 
should be taken on the same
sample, using the same experimental technique, such as infrared or
microwave spectroscopy. Using the data from different sources can lead
to conflicting results \cite{blundell05,dordevic13,tajima05}.
The experimental values for optimally doped LSCO that
we have compiled from the literature fully support this argument. 
To illustrate the point, in Table~\ref{lsco} we
list the values of superfluid density $\rho_s$ for optimally doped
LSCO from several different sources. It can be seen 
that Bo\v{z}ovi\'{c}'s film has
the highest superfluid density ($\rho_{s,max}$), and that other values
can be smaller by as much as 80 $\%$. We also note that the value of $\rho_s$
obtained in Ref.~\onlinecite{armitage11} on Bo\v{z}ovi\'{c}'s film, but using a different
experimental technique is about 30 $\%$ smaller. Similarly, 
Tajima's IR and muon spectroscopy measurements have resulted 
in superfluid densities that differ by almost a factor of three 
\cite{tajima05,comm-tajima}. Therefore one must not
compare the absolute values of superfluid density obtained using 
different experimental techniques, even when taken on the same 
sample or film. However, we point out that the measurements
in Ref.~\onlinecite{bozovic16} were performed on thousands of samples, 
grown and measured
using the same procedure, which assures that their relative values 
(i.e. their {\it doping} dependence) are
reliably extracted and are intrinsic property of these LSCO films.

\begin{table*}[t]
\centering
\begin{tabular}{|c|c|c|c|c|}

\hline

Sample type & Experimental technique & $\rho_s$ ($\times$~10$^7$ cm$^{-2}$) & $\rho_s$/$\rho_{s, max}$  & Reference  \\ \hline \hline

Bo\v{z}ovi\'{c}'s film & Mutual inductance & 6.17 & 100 $\%$   &  Ref.~\onlinecite{bozovic16} \\ \hline

Bo\v{z}ovi\'{c}'s film & THz spectroscopy & 4.53 & 73 $\%$  & Ref.~\onlinecite{armitage11}  \\ \hline

Lemberger's film & Mutual inductance & 5.97 & 97 $\%$ & Ref.~\onlinecite{lemberger11}  \\ \hline


Tajima's single crystal & IR spectroscopy & 1.17 & 19 $\%$ & Ref.~\onlinecite{tajima05} \\ \hline

Tajima's single crystal & Muon spectroscopy & 3.05 & 49 $\%$ & Ref.~\onlinecite{tajima05} \\ \hline

van der Marel's single crystal & IR spectroscopy & 1.32 & 21 $\%$ & Ref.~\onlinecite{marel21} \\ \hline

\end{tabular}

\caption{Superfluid density $\rho_s$ of optimally doped LSCO from six different sources.
The first three are from thin films, whereas the last three are from bulk single crystals.
$\rho_{s, max}$ is the value from Ref.~\onlinecite{bozovic16}. Note significant differences
in superfluid density for the same film or bulk sample, extracted using two different 
experimental techniques.}

\label{lsco}


\end{table*}


\section{Discussion}

In this section we discuss possible scenarios that 
might be able to account for the effects observed above. It has 
been know for a long time that in thin films of conventional 
superconductors (such as Sn \cite{orr86}) the so-called microscopically 
granular superconductivity might arise. These systems were
modeled as Josephson junction arrays and it was shown that quantum 
fluctuations play an important role in suppressing superconductivity 
in them \cite{chakravarty86,fisher87}. Similar ideas were also discussed 
in relation to superconductivity in the cuprates \cite{imry08}, and they
might also apply to overdoped LSCO films. Further support comes
from a recent experimental study of heavily overdoped LSCO
\cite{yangmu22}.

More recently, quantum phase fluctuations have
been argued to explain strong suppression of superfluid density in 
Bo\v{z}ovi\'{c}'s overdopded LSCO films. Schneder employed finite 
size scaling analysis \cite{schneider21} which uncovered that
suppression is consistent with a finite length limited 3D-XY
transition \cite{lsco-comm}; in some films this limiting length is set by the film 
thickness, and in others by inhomogeneities. Moreover, the analysis reveals 
a crossover from thermal to quantum critical regime as 
T$_c$~$\rightarrow$ 0, and Schneider argues that in Bo\v{z}ovi\'{c}'s 
overdoped LSCO films the suppression of superfluid density is driven
by quantum phase fluctuations.

Additional experimental support for quantum phase fluctuations'
driven suppression of superfluid density comes from the recent
terahertz spectroscopy measurements on Bo\v{z}ovi\'{c}'s LSCO films
by Mahmood {\it et al.} \cite{armitage18}. They discovered that below
T$_c$ a significant fraction of charge carriers remains uncondensed 
in a wide Drude-like peak \cite{armitage18} and argued that quantum phase
fluctuations play an important role in suppressing the superfluid density.

Based on all these findings we suggest that in thin films quantum 
confinement (i.e. reduced dimensionality) enhances
quantum phase fluctuations, making them more efficient
in preventing pair formation and reducing the superfluid density.
This might result in suppression of superfluid density
in overdoped thin films, compared with bulk samples. If this 
suggestion is correct, one might expect that similar behavior could 
also be observed in pnictide thin films. 

The final issues that we discuss here is what happens as the quantum 
critical point (T$_c$~$\rightarrow$ 0) is approached. In thin
films the superfluid density is continuously suppressed and goes 
to zero as T$_c$~$\rightarrow$ 0. Based on Fig.~\ref{fig:overdoped}
we hypothesize that in bulk samples the superfluid density will 
not be continuously suppressed to zero, but instead will experience
an abrupt drop once the critical doping is reached. Measurements
on bulk samples with such high doping levels are currently not available.

\section{Summary}

In summary, we have explicitly shown that Bo\v{z}ovi\'{c}'s overdoped LSCO films do indeed
violate Homes' law. Analyzing the existing literature data, we have unraveled
that the main reason for this violation is the stronger suppression of superfluid density
in thin films, compared to bulk single crystals. We hypothesize that in
thin films superfluid density is suppressed by quantum phase fluctuations. Our results 
have uncovered a fundamental difference between the superfluid density in
bulk samples and thin films. These findings call for measurements of superfluid
density in bulk cuprate samples close to the quantum critical point,
i.e. in the region where T$_c$~$\rightarrow$ 0, as well as on overdoped pnictide films.

{\it Note added in proof:} We only recently became aware of
two studies Refs.~(\onlinecite{brewer15,deepwell13}) reporting superfluid density in bulk
single crystals of overedoped Tl-2201. The results of these
papers are consistent with our main findings.

\section{acknowledgments}

We thank I. Bo\v{z}ovi\'{c} for useful discussions.
Work at Brookhaven National Laboratory was supported by the U.S.
Department of Energy, Office of Basic Energy Sciences, 
Division of Materials Sciences and Engineering under
Contract No. DE-SC0012704.

%
%

\begin{references}
\bibitem{poly-comm} Cuprates are highly anisotropic materials, with 
vastly different values of superfluid density along ab-plane and 
c-axis directions. In our opinion, extracting superfluid density, or any other 
transport or thermodynamic property, from ceramic, sintered, polycrystalline 
or powder samples is unreliable and untrustworthy.
We do point out, however, that saturation or even increase of superfluid
density on the overdoped side was observed in polycrystalline LSCO \cite{panagopoulos03}, 
polycrystalline 
Hg-1201 \cite{panagopoulos03}, ceramic Bi-2212 \cite{uemura93}, ceramic Bi-2223 \cite{uemura93}
and ceramic Y-123 \cite{uemura93}. The only two cuprate families in which 
a decrease in superfluid density was observed were sintered ceramic \cite{uemura93} and 
polycrystalline \cite{niedermayer93} Tl-2201, as well as polycrystalline 
Y$_{0.8}$Ca$_{0.2}$Ba$_2$(Cu$_{1-z}$Zn$_z$)$_3$O$_{7-\delta}$ 
\cite{bernhard01}. The effect was referred to as the 'boomerang' effect
\cite{bernhard95,tallon03}.
\bibitem{comm-Tcmax} The values of T$_{c,max}$ for the cuprates are: 40 K for LSCO
and 96 K for Bi2212. For the pnictides,  
T$_{c,max}$ is 31 K for BaFe$_2$(As$_{1-x}$P$_x$)$_2$ 
and 34 K for Ba$_{1-x}$K$_x$Fe$_2$As$_2$.
\bibitem{film-comm} There are numerous superconducting materials 
which display different behavior (i.e. different T$_c$ and other parameters) 
in bulk and film forms. The most extreme example is the recently discovered \cite{li19}
superconductor Nd$_{0.8}$Sr$_{0.2}$NiO$_2$ which is superconducting
in the form of films, but not in bulk \cite{li20}.
\bibitem{pnictide-comm} Saturation of superfluid density on the overdoped
side was also obsreved in polycrystalline LaO$_{1-x}$F$_x$FeAs \cite{luetkens09}.
\bibitem{comm-tajima} Similar discrepancies in superfluid density obtained
from IR and muon spectroscopy measurements were also noticed in other cuprate 
families, such as YBCO and Bi2212 \cite{tajima05}.
\bibitem{lsco-comm} It was shown in Ref.~\onlinecite{ohashi09} that
critical charge dynamics in overdoped LSCO films is of unknown type
(2D-"U").
\end{references}

\end{document}